\documentclass[iicol]{sn-jnl}

\usepackage{graphicx, multirow, amsmath, amssymb, amsfonts, amsthm, mathrsfs, textcomp, manyfoot, booktabs, algorithm, algorithmicx, algpseudocode, listings, tabularx, enumitem, float, bm, layouts, marvosym}
\usepackage[dvipsnames]{xcolor}
\usepackage[title]{appendix}
\usepackage[utf8]{inputenc}
\usepackage[T1]{fontenc}

\usepackage[nameinlink,capitalise]{cleveref}
\crefname{equation}{Eq.}{Eqs.}
\crefname{figure}{Fig.}{Figs.}
\crefname{tabular}{Table}{Tables}
\crefname{section}{Section}{Sections}
\crefname{subsection}{Section}{Sections}
\crefname{subsubsection}{Section}{Sections}
\crefname{appendix}{Appendix}{Appendices}

\usepackage[sorting=none, style=numeric-comp, natbib=true, maxbibnames=99, maxcitenames=2, giveninits]{biblatex}
\AtEveryBibitem{%
    \clearfield{issn}
    \clearfield{urlyear}
    \clearfield{urlmonth}
    \clearfield{month}
    \clearfield{pages}
    \ifentrytype{online}{}{
        \clearfield{url}
    }
}
\renewbibmacro{in:}{}

\begin{document}

\title{\bf Shear jamming and nonlinear rheology of chocolate suspensions}

\author*[1]{\fnm{Michel} \sur{Orsi}}\email{mich.orsi@gmail.com}

\author[1]{\fnm{Veeraj} \sur{Shah}}\email{veerajshah17@gmail.com}

\author[2]{\fnm{Mahesh} \sur{Padmanabhan}}\email{mahesh.padmanabhan@mdlz.com}

\author[3]{\fnm{Thomas} \sur{Curwen}}\email{thomas.curwen@mdlz.com}

\author[1]{\fnm{Jeffrey F.} \sur{Morris}}\email{morris@ccny.cuny.edu}

\affil*[1]{\orgdiv{Levich Institute and Dept. of Chemical Engineering}, \orgname{CUNY City College of New York}, \orgaddress{\city{New York}, \state{New York}, \postcode{10031} \country{USA}}}

\affil[2]{\orgname{Mondel\={e}z International}, \orgaddress{\city{East Hanover}, \state{New Jersey}, \postcode{07936} \country{USA}}}

\affil[3]{\orgname{Mondel\={e}z International}, \orgdiv{The Reading Science Centre}, \orgaddress{\city{Reading}, \state{Berkshire}, \postcode{RG6 6LA} \country{UK}}}

\maketitle

\begin{abstract}

We experimentally investigate the rheology of dark chocolate pastes in both industrially relevant pre-refined form and simplified model systems. Steady and oscillatory shear experiments reveal yielding, pronounced shear-thinning, and stress-dependent hysteresis governed by solid loading. Fitting the viscosity data with the Maron-Pierce model provides stress-dependent maximum flowable fractions $\phi_{\rm{m}}(\sigma)$, defining yield loci in the $(\phi, \sigma)$ plane. Their variation with stress quantifies the coupled roles of friction and adhesion in setting flow limits. Large-amplitude oscillatory shear tests characterize transitions from elastic to viscous behavior and identify distinct recovery pathways near jamming. Contact-stress decomposition separates hydrodynamic and frictional contributions, confirming that adhesive contact networks dominate stress transmission in pre-refined pastes. These results establish chocolate pastes as dense, adhesive suspensions whose flow is controlled by the interplay of friction and adhesion, offering quantitative benchmarks for constitutive modeling and linking chocolate processing to the broader physics of constraint rheology.

\end{abstract}

\section{Introduction}\label{sec:intro}

Dark chocolate is typically processed as a dense suspension in which solid sugar and cocoa solid particles are dispersed in molten cocoa butter. Sugar is usually added as crystalline sucrose, with initial particle sizes typically several hundreds of microns. These particles are reduced in size during processing to achieve the desired texture and mouthfeel of the final product. The first stage, known as \emph{pre-refining}, involves mixing the confectionery ingredients, followed by mechanical size reduction using cylindrical rollers. This produces a paste that is then further refined -- typically with a five-roll refiner (see Chapter~7 of \citet{beckett2017}, and \citet{lipkin2025}) -- to reduce the particle size below 30$\,\mu$m. The rheology of the paste entering this refining step, i.e., immediately after pre-refining, strongly influences the processability and is the primary focus of this study. These pastes are formulated without emulsifiers like lecithin or PGPR, which are only added to chocolate in the subsequent conching phase of the chocolate making process.

A number of confections involve complex fluid rheology. For example, complementary to our suspension-focused study, \citet{weir2016} show that caramels behave as emulsion-filled protein gels whose small amplitude oscillatory shear (SAOS) spectra obey time–composition superposition, yielding master curves controlled by protein-network connectivity.

Chocolate pastes pose a rheological challenge because of their polydisperse, non-spherical particulate composition. The dominant solid component, sucrose, initially crystallizes in a monoclinic form. During pre-refining, the sucrose undergoes fracturing and breakage, producing irregularly-shaped particles \cite{hartge2025,lipkin2025}. In dark chocolate formulations, cocoa solids form the second major particulate phase; these are produced by grinding cocoa nibs and typically appear as smaller granules or platelets \cite{tan2017,do2011}. The suspending phase is cocoa butter, which remains liquid under the temperature conditions relevant to this work.

Given the broad particle size distribution and anisotropic shapes, chocolate paste suspension behavior cannot be adequately described by volume fraction alone. A more informative approach is to consider the volume fraction relative to the rheologically defined maximum flowable fraction $\phi_{\rm m}(\sigma)$, which denotes the solid loading at which the suspension's viscosity diverges under an applied stress $\sigma$; this divergence is often empirically described by the Maron-Pierce model \cite{maron1956}. At high stresses, $\phi_{\rm m}$ approaches the jamming threshold, $\phi_{\rm m} \rightarrow \phi_{\rm J}$; at lower stresses, it defines a stress-dependent yield point, i.e., $\phi_{\rm m}(\sigma)$ can be inverted to find the yield stress $\sigma_{\rm y}(\phi)$ \cite{orsi2025,richards2020}. In related work, \citet{blanco2019} interpret conching as ‘jamming engineering', showing that aggregate breakup and dispersant-induced lubrication raise $\phi_{\rm J}$ and reduce yield and high-shear viscosities in model chocolates.

To characterize the material across its rheological range, we adopt a two-part experimental strategy. First, to probe the flow regime, we dilute the paste to solid loadings below those encountered in production and perform steady rotational rheometry over a range of concentrations. This allows extrapolation to the viscosity divergence, and thus determination of $\phi_{\rm m}$ (and of $\phi_{\rm J}$ at high stress), providing insight into the strength and origin of interparticle interactions. Second, to explore the jammed regime at $\phi \gtrsim \phi_{\rm J}$, we perform large-amplitude oscillatory shear (LAOS) measurements over a range of strain amplitudes and solid loadings. These measurements provide insight to how contact networks develop and evolve, and how these effects depend on the paste composition in terms of the mix of ingredients. Our results show that chocolate pastes exhibit both yielding and shear jamming. We distinguish these behaviors and provide clarification of their respective physical origins.

In the following section, we detail the experimental methodology and sample preparation. Then, we present the rheological results and interpret them in light of the underlying particle-scale mechanisms. Finally, we discuss conclusions and broader implications of the work.

\section{Experimental protocols}\label{sec:experiments}

\subsection{Rheometry}\label{sec:rheometry}

All rheological measurements are performed using an ARES-G2 rheometer (TA Instruments) equipped with a 50$\,\rm{mm}$ parallel-plate geometry maintained at 40{\textdegree}C. This geometry is chosen because, in such torsional flow, there is no or weak shear-induced particle migration \cite{merhi2005,chow1994}. The drawback is that the shear rate varies across the gap, increasing from $\dot{\gamma} = 0$ at the center to the rim value of $\dot{\gamma} = \dot{\gamma}_{R} = \Omega R / h$, with $\Omega$, $R$, and $h$ being the angular velocity, the plate radius, and the gap height, respectively. This variation is taken into account by applying the well-known Mooney-Rabinowitsch viscosity correction:
\begin{equation}
    \eta_{\rm{susp}} = \eta_{\rm{app}} \left[1 + \dfrac{1}{4}\dfrac{\rm{d} \ln\left(\eta_{\rm{app}}\right)}{\rm{d}\ln\left(\dot{\gamma}_{R}\right)}\right]\,,
\end{equation}
where $\eta_{\rm{app}} = 2\Gamma / \left(\pi R^3 \dot{\gamma}_{R}\right)$ is the apparent viscosity measured by the rheometer, $\Gamma$ being the applied torque.

\begin{figure*}[!t]
    \includegraphics[height=0.183\linewidth]{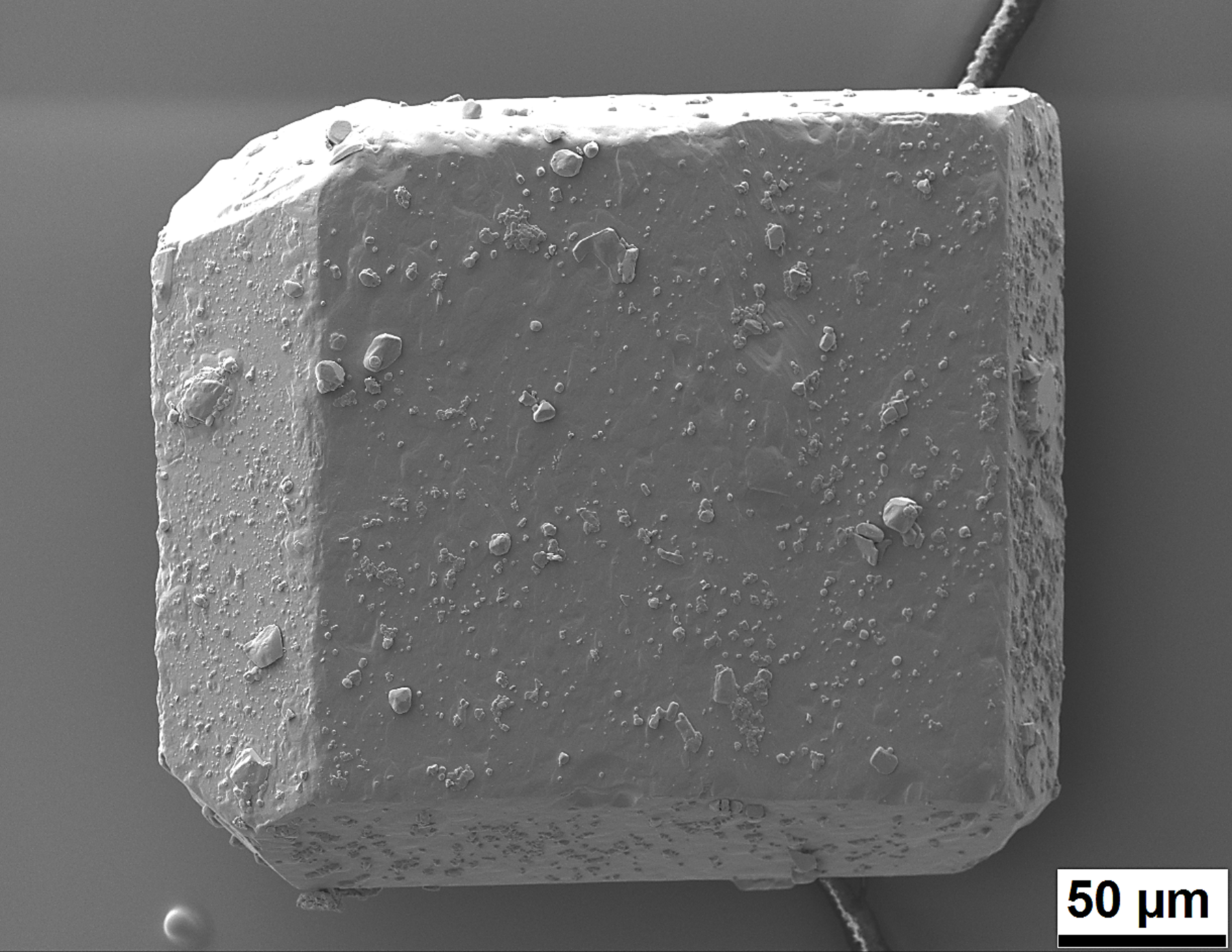}\hspace{2pt}\includegraphics[height=0.183\linewidth]{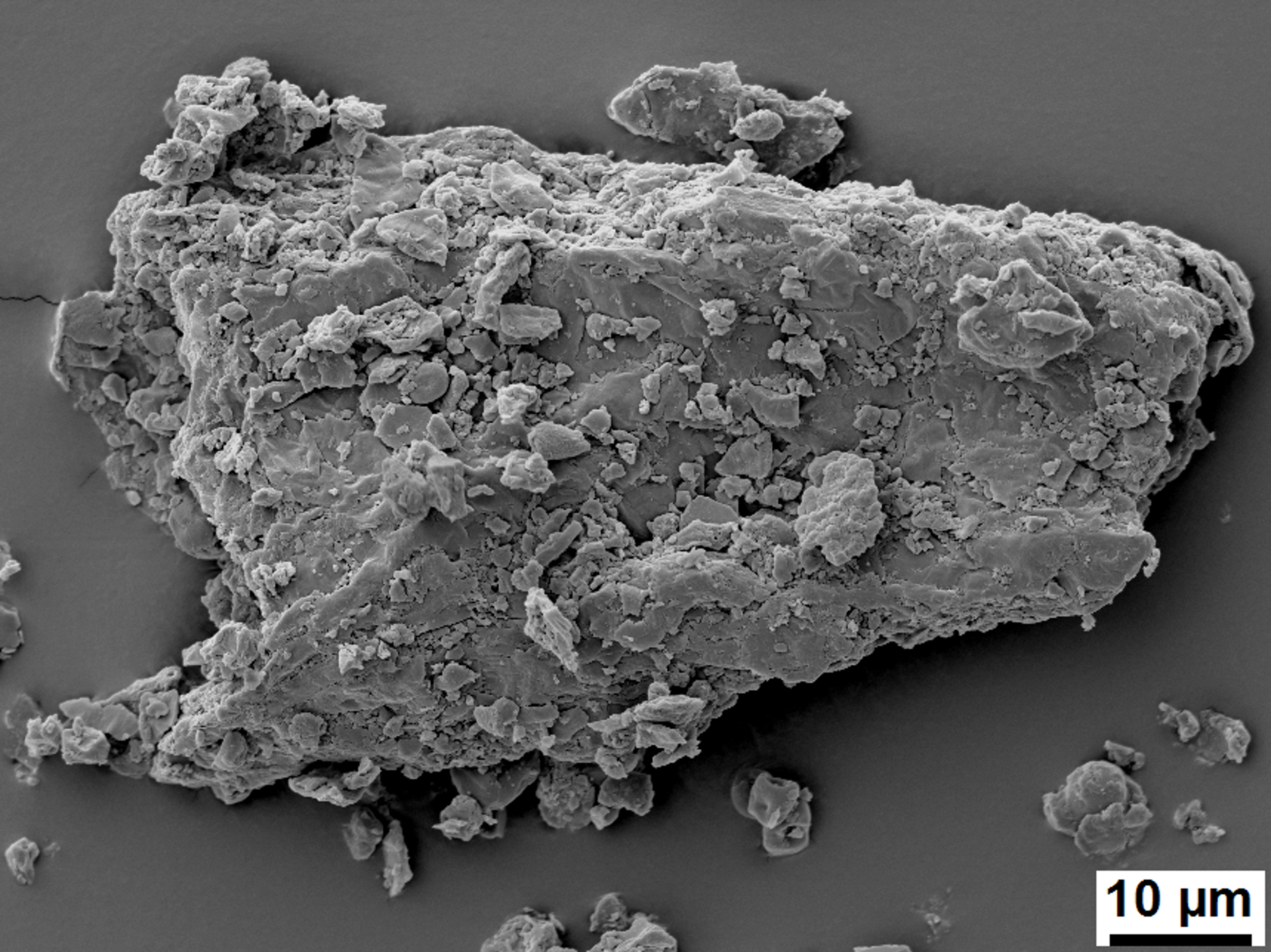}\hspace{2pt}\includegraphics[height=0.183\linewidth]{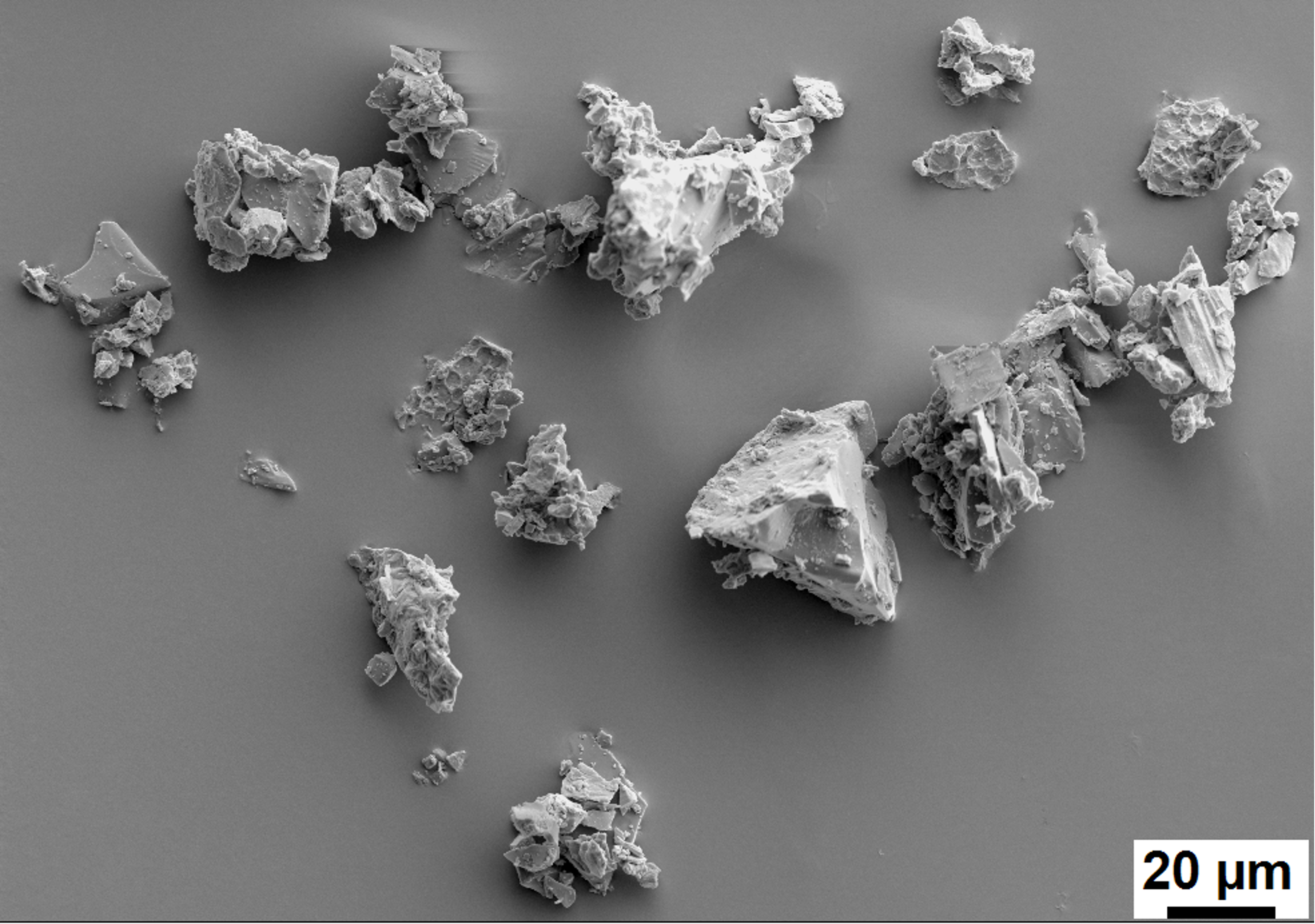}\hspace{2pt}\includegraphics[height=0.183\linewidth]{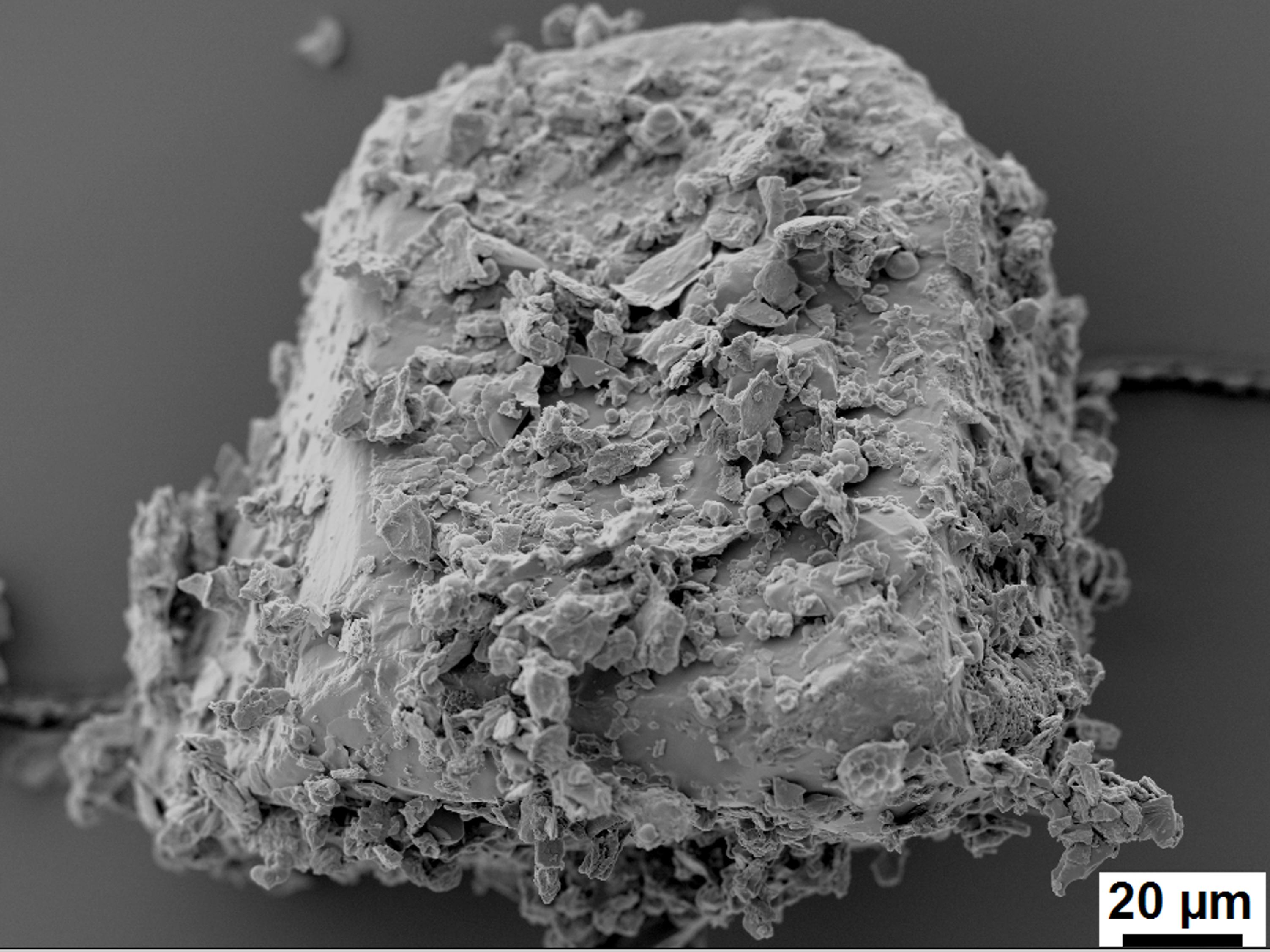}
    \vspace{-10pt}
    \caption{Electron microscopy images of (from left to right) sucrose particle (one large with a few fragments), cocoa solid (one large and many smaller cocoa particles), pre-refined dark chocolate paste (more irregular sugar crystals mixed with cocoa particles that get nearly as big as the sugar but are also much smaller), and the dark model paste (one large sugar crystal decorated in cocoa particles).}
    \label{fig:particle_images}
\end{figure*}

To avoid wall slip, 220-grit sandpaper (Dura-Gold) is affixed to both plates; the gap is then zeroed; the sample is loaded until the lower plate is completely covered, and the surface is leveled with a spatula to a thickness slightly exceeding the target gap; finally, the upper plate is lowered until it fully contacts the sample, at which point the actual gap height is recorded. Accurate viscosity measurements require a gap at least $\approx$20 times the particle diameter \citep{zarraga2000}. The pre-refined samples are polydisperse with a characteristic size $D_{90} = 160\,\mu\rm{m}$ ($D_{90}$ denotes the particle size below which $90\%$ by volume of the particles fall; see \cref{fig:particle_images} for microscopy images of the particles involved, and \cref{fig:PSDs} for the particle size distributions of the samples), and to roughly agree with this criterion the measurement gap is maintained between 2 and 3$\,\rm{mm}$. The characteristic size for the sucrose particles is $D_{90} = 400\,\mu\rm{m}$; hence, in the case of non-pre-refined model pastes, a gap of 4 to 5$\,\rm{mm}$ is employed. Multiple measurements with different gap height have been performed to check that the obtained values are not gap-dependent \cite{yoshimura1988}.

Given the typical particle size $a$ involved, the shear stresses $\sigma$ investigated, and the gap size $h$ used, the Reynolds number $Re = \rho \sigma h^2 / \eta^2 < 0.1$ is small enough for inertial effects to be negligible, and the P\'{e}clet number $Pe = 6\pi\sigma a^3 / (k_{B}T) > 10^9$ is large enough to neglect Brownian effects, where $k_B$ is the Boltzmann constant and the absolute temperature is $T$.

\subsection{Materials and sample preparation}\label{sec:samples}

\subsubsection{Pre-refined samples}\label{sec:prerefinedsamples}

The dark chocolate pastes studied here are composed of cocoa butter (referred to as ``fat'' in the following, for brevity), cocoa solids, and crystalline sucrose. These undergo a pre-refining step that reduces the particle size by passage between closely spaced metal rollers. The initial samples have, as noted, a particle size distribution with $D_{90} = 160\,\mu\rm{m}$ and contain $20\,\rm{wt}\%$ fat.

\begin{table}[!b]
    \centering
    \caption{Composition of pre-refined $20\,\rm{wt}\%$ fat dark chocolate paste.}
    \label{tab:recipes}
    \begin{tabular}{|c|c|c|} \hline
    sucrose & cocoa liquor & cocoa butter \\ \hline
    64.5 \rm{wt}\% & 34.2 \rm{wt}\% & 1.3 \rm{wt}\% \\ \hline
    \end{tabular}
\end{table}

The base paste is prepared by Mondel\={e}z in Reading (UK), with the composition shown in \cref{tab:recipes}. Notably, $54.9\,\rm{wt}\%$ of the cocoa liquor is fat, of which approximately $20\,\rm{wt}\%$ is trapped within the porous cocoa solid particles \citep{do2011}, contributing to the total cocoa solids volume fraction. Volume fractions are calculated using densities measured at 40{\textdegree}C: $\rho_{\text{sucrose}} = 1.59\,\rm{g}\,\rm{cm}^{-3}$ for sucrose, $\rho_{\text{fat}} = 0.89\,\rm{g}\,\rm{cm}^{-3}$ for cocoa butter, and $\rho_{\text{cocoa}} = 1.37\,\rm{g}\,\rm{cm}^{-3}$ for cocoa solids (including trapped fat). The cocoa butter viscosity measured in this work is $\eta_{\rm fat} = 0.046\,\rm{Pa}\,\rm{s}$ at 40{\textdegree}C, consistent with literature values \citep{landfeld2000,mishra2021}.

To prepare samples with different fat contents, both the base paste and additional fat are warmed to 40{\textdegree}C. A mass $\Delta M = {M_0} (w - w_0)/(1-w)$ of liquid fat is added, where $w$ is the target fat mass fraction, and $w_0$ and $M_0$ are the initial fat fraction and mass of the base paste, respectively. Blending is achieved by gentle stirring for several minutes in a stand mixer (Kenwood).  Samples are stored at 40{\textdegree}C until measurement.

\subsubsection{Model samples}

Model samples are prepared from individual ingredients in as-received condition, i.e., materials which have not been subject to pre-refining. Here, the composition is defined by the solid volume fraction rather than the fat weight fraction. Additionally, samples consisting only of fat and cocoa powder are prepared; these are termed ``liquor'' pastes, \textit{chocolate liquor} being the term historically used to indicate the suspension produced when roasted cocoa nibs are ground, containing both cocoa solids and cocoa butter.

The dark chocolate model samples are prepared by first making a $20\,\rm{wt}\%$ fat paste following the recipe in \cref{tab:recipes} and then diluting it by adding cocoa butter to obtain a range of computed solid volume fractions spanning from $\phi = 0.47$ to $\phi = 0.71$ in steps of 0.02. The cocoa liquor samples are prepared by simply using the densities given in the previous section.

The main distinction between pre-refined and model samples is that the latter are not subjected to pre-refining -- as a result, particles are larger in the model sample (see next section).

\subsection{Particle populations}\label{sec:PSDs}

\begin{figure}[!b]
    \centering
    \includegraphics[width=\linewidth]{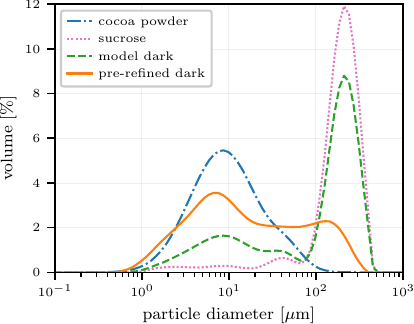}
    \caption{Particle size distributions of materials used in this work: cocoa powder, sucrose, as well as model and pre-refined dark chocolate pastes.}
    \label{fig:PSDs}
\end{figure}

\cref{fig:particle_images} shows electron microscopy images of sucrose, cocoa solids, and pre-refined and dark chocolate pastes. The particle size distributions (PSDs) -- measured with a Malvern Mastersizer -- of cocoa powder, sucrose, and model and pre-refined dark chocolate pastes are shown in \cref{fig:PSDs}. Samples for particle size measurement were prepared by dispersing the powders or diluting the pastes in Akomed with ultrasound to ensure that any aggregates were broken up. The PSD of the liquor paste corresponds to that of cocoa powder, while the PSD of the model dark chocolate paste corresponds to a weighted sum of the PSDs of cocoa powder and sucrose.

It should be noted that the PSD of the pre-refined dark chocolate paste is significantly different from its model counterpart. Pre-refining reduces the size of the majority of the sucrose particles, creating a new population slightly smaller than the cocoa solids.

\subsection{Steady simple shear flow}\label{sec:simpleshear}

Each experiment begins with a ``pre-shear'' step, in which the sample is sheared at $\sigma = 50\,\rm{Pa}$ for 10 minutes. This procedure serves two purposes: (i) it erases any structural memory introduced during sample loading between the rheometer plates, and (ii) it promotes a more homogeneous, reproducible microstructure. The chosen duration also ensures uniform temperature throughout the sample.

During this pre-shear, the initial transient response -- corresponding to the first $100\, \rm{s}$ and equivalent to a strain exceeding two in all cases -- is discarded. Only the subsequent steady-state data are averaged and retained for analysis.

\begin{figure}[!t]
    \centering
    \includegraphics[width=\linewidth]{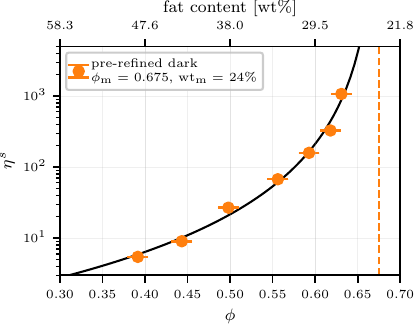}
    \caption{Pre-refined samples. Relative viscosity as a function of both volume fraction (bottom axis) and fat content (top axis) at $\sigma = 50\,\rm{Pa}$. The Maron-Pierce fit is shown as the solid curve, with the corresponding maximum volume fraction $\phi_{\rm{m}} = 0.675$, which corresponds to a maximum fat content $\rm{wt}_{\rm{m}} = 24\%$. The optimal fitting parameters are found to be $\alpha = 0.57$ and $\beta = 2.7$. Error bars denote standard deviation ($\pm\,\rm{std}(\eta^s)$) across independent loadings. Representative raw signals at are shown in \cref{fig:reproducibility_PreRef}.}
    \label{fig:etaS_vs_phi_stress50Pa_PreRef}
\end{figure}

\begin{figure}[!t]
    \centering
    \includegraphics[width=\linewidth]{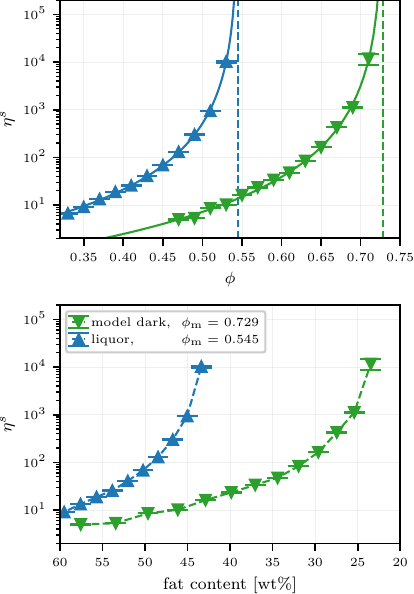}
    \caption{Model samples. Relative viscosity as a function of (top) volume fraction and (bottom) fat content at $\sigma = 50\,\rm{Pa}$. The Maron-Pierce fit curves are shown as solid lines in the top panel, while the dashed lines in the bottom panel are a visual guide. Estimated maximum volume fractions $\phi_{\rm{m}}$ are indicated in the legend. For the model dark paste, $\alpha = 0.24$ and $\beta = 2.9$; for the liquor, $\alpha = 0.58$ and $\beta = 2.7$. Error bars denote standard deviation ($\pm\,\rm{std}(\eta^s)$) across independent loadings.}
    \label{fig:etaS_vs_phi_stress50Pa_Model}
\end{figure}

Three independent loadings were performed for each sample and composition; we report the mean and standard deviation (SD) of the steady viscosity across loadings. Error bars in \cref{fig:etaS_vs_phi_stress50Pa_PreRef,fig:etaS_vs_phi_stress50Pa_Model} reflect these SDs. Representative raw signals for the pre-refined paste at $\sigma=50$\,Pa and 28, 32, and 40\,wt\% are provided in \cref{fig:reproducibility_PreRef} in \cref{app:reproducibility}. For all compositions, the SD of steady-state viscosity across loadings is below $10\%$ of the mean value, confirming reproducibility.

The relative viscosities, measured at various solid volume fractions $\phi$ (determined from fat content as described in \cref{sec:prerefinedsamples}), are fitted to the Maron-Pierce relation:
\begin{equation}\label{eq:MaronPierce}
    \eta^{s}\left(\phi, \sigma\right) = \frac{\alpha\left(\sigma\right)}{\left[ 1 - \phi / \phi_{\rm m}(\sigma) \right]^{\beta}}\,,
\end{equation}
where $\phi_{\rm m}(\sigma)$ is the maximum flowable volume fraction under the imposed stress $\sigma$. $\alpha(\sigma)$ reflects flow conditions, while $\beta$ depends on material properties. Here, both are treated as free fitting parameters; for spheres, $\alpha \approx 1$ and $\beta \approx 2$ \citep{maron1956,krieger1959}.

\cref{fig:etaS_vs_phi_stress50Pa_PreRef} shows the steady-state relative viscosity of the pre-refined pastes, i.e., the measured viscosity normalized by the viscosity of molten cocoa butter at the experimental temperature, as a function of volume fraction and fat content. The Maron-Pierce fit (solid curve) matches the data well and diverges at $\phi_{\rm m}(\sigma=50\,\rm{Pa}) \doteq 0.675$, corresponding to $\approx 24\, \rm{\rm{wt}\%}$ fat content.

Results for the model samples are shown in \cref{fig:etaS_vs_phi_stress50Pa_Model}, again plotted as a function of volume fraction and fat content. Note that the fit is only expected to fit relatively near jamming, as the leading coefficient $\alpha < 1$ is too small for $\eta^s$ to tend toward unity as $\phi\rightarrow 0$.

Comparing the three systems at fixed $\phi$ reveals a consistent hierarchy: the model liquor paste exhibits the highest relative viscosity, the model dark chocolate paste exhibits the lowest relative viscosity, and the pre-refined dark chocolate paste falls between these two.

This ordering correlates with PSDs of the materials. Narrow PSDs generally yield smaller values of $\phi_{\rm m}$ or $\phi_{\rm J}$, and thus a larger relative viscosity at a given $\phi$ \cite{pednekar2018}. The liquor paste has a narrow PSD of small cocoa solids; the model dark paste has a broad PSD mixing small cocoa solids with large, unrefined sucrose particles; and the pre-refined dark paste distribution is also broad but less so because the sucrose particles are generally smaller than those in the model paste.

\cref{tab:AlphasBetas} shows the fitting parameters $\phi_{\rm m}$, $\alpha$, and $\beta$ obtained for the three pastes. The deviations of the viscosity data for the pre-refined pastes from the Maron-Pierce model are reflected in larger uncertainties of the fitting parameters for this system. Nevertheless, it can be noticed that the pre-refined dark chocolate and liquor pastes show very similar values for both $\alpha$ and $\beta$, but quite different 
maximum packing fractions.

\begin{table}[!t]
    \centering
    \caption{Fitting parameters for the steady simple shear flow data at $50\,\rm{Pa}$.}
    \label{tab:AlphasBetas}
    \begin{tabular}{|c|c|c|c|} \hline
    & pre-ref. dark & model dark & liquor \\ \hline
    $\phi_{\rm m}$    & $0.675 \pm 0.02$ & $0.729 \pm 0.002$ & $0.545 \pm 0.001$ \\ \hline
    $\alpha$ & $0.57  \pm 0.30$ & $0.24  \pm 0.03$  & $0.58  \pm 0.04$  \\ \hline
    $\beta$  & $2.7   \pm 0.57$ & $2.9   \pm 0.10$  & $2.7   \pm 0.05$  \\ \hline
    \end{tabular}
\end{table}

\paragraph{On the Maron-Pierce exponent}

Across all three chocolate pastes, fitting \cref{eq:MaronPierce} at $\sigma=50\,\rm{Pa}$ returns $\beta\simeq 2.7$--$2.9$ (Table~\ref{tab:AlphasBetas}), larger than the $\beta\approx 2$ often quoted for nearly spherical, non-Brownian hard-sphere suspensions in Newtonian media. In this work, $\beta$ is treated as an \emph{empirical} fit parameter that quantifies how sharply $\eta^s$ diverges as $\phi\to\phi_{\rm m}(\sigma)$; our data do not isolate a unique microscopic origin for the larger exponent. More generally, the steepness of the divergence is not universal: in divergence forms such as Krieger-Dougherty, the effective exponent is linked to the intrinsic viscosity and a packing limit, and thus can increase when particle shape and/or interactions increase the intrinsic viscosity and the number of constraints per particle \citep{krieger1959,wildemuth1984}. Consistent with this, concentrated suspensions of non-spherical particles (including spheroids/platelets) exhibit strong shape- and aspect-ratio-dependence of viscosity and divergence behavior, implying a steeper approach to the flow limit than in smooth-sphere suspensions \citep{faroughi2017,pandare2026}.

With that context, we interpret $\beta>2$ here as consistent with (and potentially arising from) several chocolate-specific features that could steepen the divergence near $\phi_{\rm m}(\sigma)$: (i) adhesive, friction-bearing contacts, yielding contact networks that carry load and sharpen the growth of viscosity as $\phi\to\phi_{\rm m}(\sigma)$; (ii) deviations from sphericity and surface roughness (angular sucrose, irregular cocoa) that increase kinematic constraints per contact (sliding and rolling), effectively increasing the crowding penalty relative to sphere suspensions; and (iii) broad/bimodal particle-size distributions that raise packing efficiency (increase $\phi_{\rm m}$) via void filling, while still allowing percolated contact networks at a given global $\phi$ \citep{liu2021,laine2006}. Taken together, these factors provide a plausible rationale for the empirically observed $\beta\simeq 3$ in these dense, contact-dominated pastes.

\subsection{Shear-rate sweeps}\label{sec:shearrate}

\begin{figure}[!t]
    \centering
    \includegraphics[width=\linewidth]{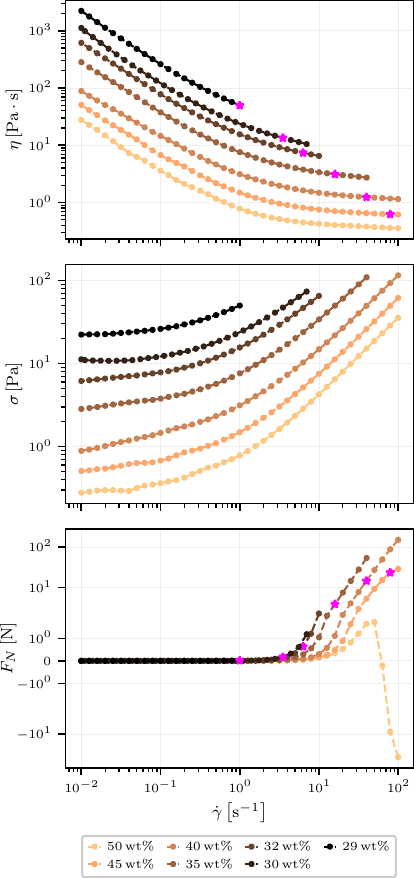}
    \caption{Shear-rate sweeps for pre-refined pastes. (top) viscosity, (center) shear stress, and (bottom) normal force vs. shear rate for different fat contents. Lines are a visual guide. Magenta stars ({\color{magenta}$\star$}) in top and bottom panels indicate points at $\approx 50\,\rm{Pa}$ as a link to \cref{fig:etaS_vs_phi_stress50Pa_PreRef,fig:etaS_vs_phi_stress50Pa_Model}.}
    \label{fig:RateSweeps_PreRef}
\end{figure}

After a 10-minute pre-shear at $\sigma = 50\,\rm{Pa}$, stepwise shear-rate sweeps are performed from high to low rates. The maximum rate is set according to the value measured during pre-shear, which depends on fat content; the minimum is fixed at $\dot{\gamma} = 10^{-2}\,\rm{s}^{-1}$. This choice avoids edge fracture in low-fat samples, a phenomenon linked to large flow-induced normal stresses \cite{keentok1999}. Performing a descending rate sweep after a high-stress pre-shear lets the samples start from a well-defined, fully broken-down state and then characterizes the recovery of the structure as stress is decreased -- an approach widely used to avoid slip/artifacts and to obtain reproducible flow curves (e.g., \citet{ewoldt2015} who run steady-state flow sweeps from high to low rates; and \citet{chen2021}, who explicitly use the ascending leg to disrupt structure and the descending leg to report the flow behavior). At each step, steady-state data are collected, with longer sampling times at lower shear rates to ensure comparable accumulated strain. The temperature is held at 40{\textdegree}C.

This protocol is applied only to pre-refined pastes. As seen in Fig.~\ref{fig:RateSweeps_PreRef}, across the full range of shear rates, the material exhibits pronounced shear-thinning behavior and clear yielding for all fat contents, typical of adhesive suspensions \cite{papadopoulou2020,gillissen2020b,orsi2025}. The normal force depends strongly on fat content: at high shear rates, the highest-fat sample shows negative normal force, while all the other samples show positive values increasing with volume fraction; at low shear rates, all samples show near-zero or positive normal force. On decreasing‐rate sweeps the system remains on (or returns to) a contact-rich, dilatant branch so parallel-plate thrust is dominated by positive dilatancy and confinement, which suppresses negative bulk contributions to $F_N$ \cite{chu2014,brown2014,brown2012,han2019}. Note that in parallel-plate rheometry the measured normal force is a total axial thrust, reflecting bulk normal stresses and boundary confinement; on descending rate sweeps this favors non-negative $F_N$ even when a negative bulk contribution may exist \cite{yoshimura1988,zarraga2001a}.

\subsection{Shear-stress sweeps}\label{sec:shearstress}

Following the same pre-shear step at $\sigma = 50\,\rm{Pa}$, stepwise shear-stress sweeps are conducted by increasing the applied stress from $10^{-2}$ to $10^{2}\,\rm{Pa}$. Viscosity and shear rate are continuously monitored during each step. To prevent excessively long measurement times below the yield stress and to avoid sample ejection at high shear rates, the protocol imposes two termination criteria: the step ends if viscosity reaches $10^4\,\rm{Pa}\,\rm{s}$ or shear rate exceeds $\dot{\gamma} = 200\,\rm{s}^{-1}$. Sampling durations are thus longer at low stresses, allowing careful observation of the yielding process. Temperature is maintained at 40{\textdegree}C throughout.

\begin{figure}[!t]
    \centering
    \includegraphics[width=\linewidth]{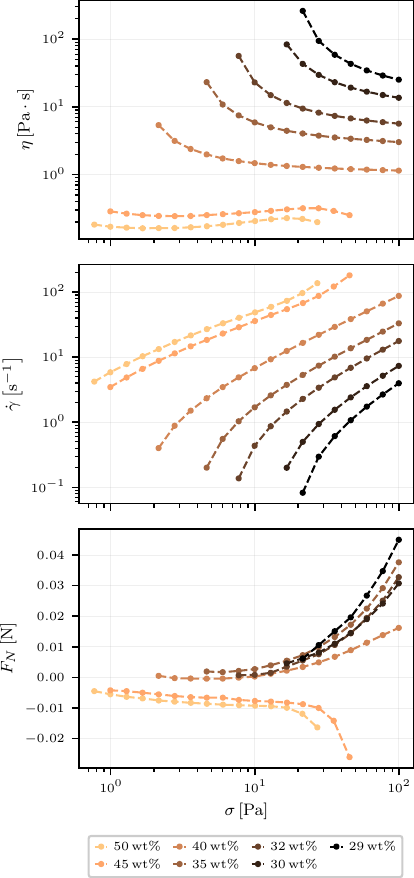}
    \caption{Shear-stress sweeps for pre-refined pastes. (top) viscosity, (center) shear rate, and (bottom) normal force vs. shear stress, different fat contents. Lines are a visual guide.}
    \label{fig:StressSweeps_PreRef}
\end{figure}

\begin{figure*}[!t]
    \centering
    \includegraphics[width=\linewidth]{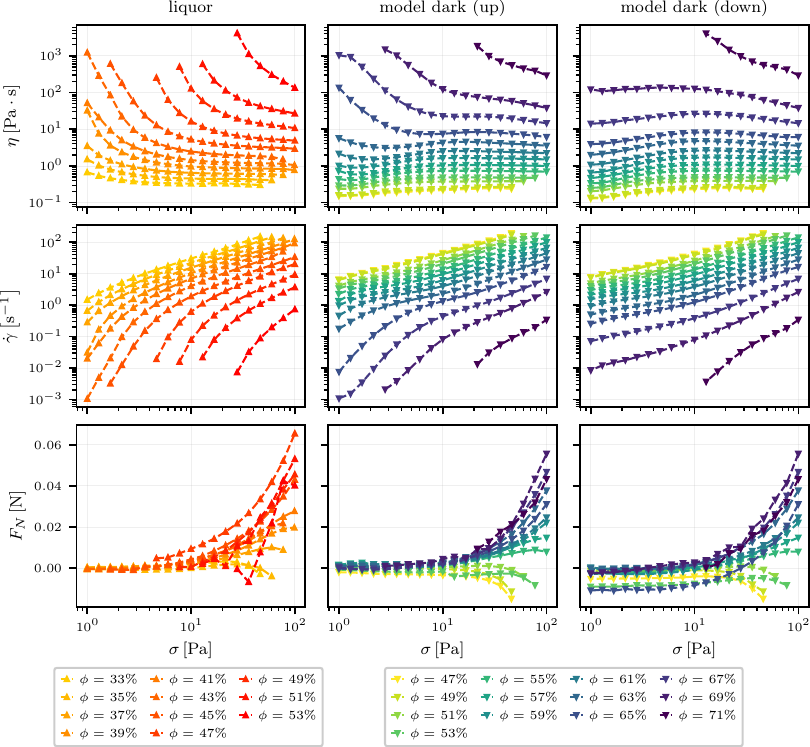}
    \caption{Shear-stress sweeps for model samples. (top) viscosity, (center) shear rate, and (bottom) normal force vs. increasing (left and center columns) and decreasing (right column) shear stress for different volume fractions. Lines are a visual guide.}
    \label{fig:StressSweeps_Model}
\end{figure*}

This procedure is applied to both pre-refined and model pastes. Data for pre-refined samples appear in \cref{fig:StressSweeps_PreRef}. Except for the two samples with the highest fat contents ($45$ and $50\,\rm{wt}\%$), results exhibit clear yield stress behavior accompanied by shear thinning across the stress range. Notably, samples with lower fat content display more pronounced shear thinning at elevated stresses. The normal force, positive for the low-fat samples, decreases with increasing fat content and changes sign beyond approximately $45\,\rm{wt}\%$, indicating a transition in microstructural response. Because the shear-rate and shear-stress sweeps protocols traverse different branches of a hysteretic flow curve -- i.e., the measured response depends on sweep direction due to dilatancy/confinement and contact rebuilding -- and sample slightly different $(\sigma,\dot{\gamma})$ windows, the stress-controlled sweep can drive high-fat pastes into a high-$\dot{\gamma}$, lubrication-dominated state with negative bulk normal stress, whereas the descending rate sweep allows frictional contacts and edge/dilatancy contributions to rebuild, masking negative $F_N$.

Although not shown here, we verified that stress- and rate-controlled pre-refined $\eta(\dot{\gamma})$ curves are in agreement, with deviations only at the two highest fat contents, where the stress-controlled protocol accesses a lower-viscosity, lubrication-dominated branch at high shear rates.

We next consider the model pastes, in an effort to determine the influence of the components in the pre-refined paste under equivalent total solid loadings, $\phi$. For the model pastes, as shown in \cref{fig:StressSweeps_Model}, both cocoa liquor and dark chocolate formulations exhibit yield stress behavior. However, in the dark chocolate paste, yielding emerges only at the higher volume fractions studied. Depending on $\phi$, dark chocolate pastes can display either shear thinning or shear thickening, while liquor pastes consistently shear thin. This indicates that the sucrose fraction is needed for shear-thickening, although this behavior may result from interaction of cocoa powder and sucrose.

Downward stress sweeps, from $\sigma = 10^{2}$ to $1\,\rm{Pa}$, are performed to probe for hysteresis effects. The results broadly mirror the upward sweeps; however, some high-$\phi$ samples lack a clear yield point in the decreasing sequence. This hysteresis evidences structural evolution under shear -- the continual formation and opening of particle contacts -- which modulates flow resistance. The effect is most pronounced near the yield stress and at elevated solids loadings, where particle networks are mechanically more robust. This behavior underscores the coupling between microstructure and flow in chocolate pastes.

\paragraph{Constraint rheology}

\begin{figure}[!t]
    \centering
    \includegraphics[width=\linewidth]{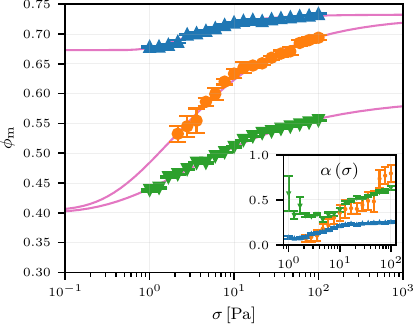}
    \caption{Maximum flowable volume fraction vs.\ shear stress, obtained by fitting data to the Maron–Pierce model (\cref{eq:MaronPierce}): ({\color{orange}$\bullet$}) pre-refined dark, ({\color{NavyBlue}$\blacktriangle$}) model dark, and ({\color{ForestGreen}$\blacktriangledown$}) liquor pastes. The inset shows the values of the fitting parameter $\alpha$ as a function of stress. The solid pink lines ({\color{CarnationPink}\Flatsteel}) are obtained by fitting the experimental results with the model proposed by \citet{richards2020} (\cref{eq:richards}), whose fitting parameters are shown in \cref{tab:RichardsParams}. In the low-stress limit: $\phi_{\rm{m}}(\sigma \rightarrow 0) = \phi_{\rm{alp}} $. Model dark shows a higher $\phi_{\rm{alp}}$ due to bimodal packing/void filling, consistent with a loose adhesive network but improved geometric packing.}
    \label{fig:phim_vs_sigma}
\end{figure}

Finally, relative viscosity data are reorganized as a function of volume fraction at various shear stresses and fitted to the Maron-Pierce model (\cref{eq:MaronPierce}) to extract -- where there is enough data -- estimates of the maximum flowable volume fraction, $\phi_{\rm m}$, as a function of applied stress. The exponent $\beta$ is fixed to values obtained from steady shear measurements at $\sigma = 50\,\rm{Pa}$ (see \cref{fig:etaS_vs_phi_stress50Pa_PreRef,fig:etaS_vs_phi_stress50Pa_Model} and \cref{tab:AlphasBetas}). The results for all three systems are displayed in \cref{fig:phim_vs_sigma}.

The signatures observed -- increase of $F_N>0$ with shear rate under rate control, contact- vs. lubrication-dominated branch selection under stress control, and the increase of the maximum fraction $\phi_{\rm{m}}$ in the low-stress limit -- are consistent with the emergence of direct, friction-bearing contacts whose constraints are progressively released with increasing stress \citep{boyer2011,bonn2017}.

The values of $\phi_{\rm m}$ at $\sigma=50\,\rm{Pa}$ closely match those determined in \cref{sec:simpleshear}, validating the approach. The $\phi_{\rm m}(\sigma)$ curves effectively define the yield locus for each system: at fixed $\phi_{\rm m}$, increasing applied stress drives flow. But this line can also be approached at fixed stress $\sigma$: in this case, increasing the volume fraction up to $\phi_{\rm m}$ takes the system into the unyielded state. Hence, this delineation partitions the $(\phi, \sigma)$ parameter space into static and flowing regimes -- a characteristic common to adhesive \cite{richards2020,orsi2025}, and soft \cite{gilbert2022}, faceted \cite{blanc2018,hafid2016}, cubic \cite{dambrosio2025}, and polygonal \cite{dambrosio2023} particle suspensions, where rolling friction imposes an additional constraint on particle motion \cite{singh2020,singh2022}.

\citet{orsi2025} showed that adhesion causes a strong variation of $\phi_{\rm m}(\sigma)$ in frictional suspensions, with the variation becoming smaller the lower the friction coefficient. As a consequence, the pronounced variation of $\phi_{\rm m}$ over two decades in stress for the pre-refined dark chocolate paste suggests the presence of strongly frictional contacts. Model dark and liquor pastes exhibit smaller but measurable stress-dependence of $\phi_{\rm m}$. Notably, $\phi_{\rm m}$ remains higher for model dark pastes compared to liquor pastes across the stress range studied, while the pre-refined paste transitions from liquor-like behavior at low stresses toward model dark behavior at high stresses. All three systems are expected to approach their adhesionless limits at large $\sigma$, although this regime was not fully reached within the stress range investigated.

The inset of \cref{fig:phim_vs_sigma} shows the values of the fitting parameter $\alpha$ as a function of shear stress: it can be noticed that its values are considerably below one and increase with stress. In this case, the values of $\alpha$ for the pre-refined paste at low stresses are closer to the values for the model dark paste; at high stresses, they become closer to the values for the liquor paste.

Finally, the results are fitted with the model proposed by \citet{richards2020}, which is based on the constraint rheology approach introduced by \citet{guy2018} and on the phenomenological model by \citet{wyart2014}, and reads:
\begin{equation}\label{eq:richards}
    \phi_{\rm m} = \phi_{\mu} + \left(\phi_{\rm{alp}}-\phi_{\mu}\right)\, a\left(\sigma\right)\,,
\end{equation}
where $\phi_{\mu}$ is the jamming volume fraction of an adhesionless suspension whose value depends on the solid friction coefficient and is well documented \cite{gallier2014,chevremont2019,mari2014,lobry2019,peters2016,arshad2021,lemaire2023}, $\phi_{\rm{alp}}$ (``alp'': adhesive loose packing) is the maximum flowable volume fraction at $\sigma \rightarrow 0$ for a suspension where all contacts are adhesive and whose value -- following \citet{richards2020} -- can be taken as the loose packing for adhesive suspensions for which numerical studies have proposed  estimates between 0.15 and 0.51 \cite{yang2000,yang2003,curk2016}, and $a(\sigma)$ is the variation of the fraction of adhesive contacts with shear stress whose expression has been proposed by \citet{richards2020} and reads:
\begin{equation}
    a\left(\sigma\right) = 1 - \exp \left[-\left(\dfrac{\sigma_a}{\sigma}\right)^{\kappa}\right]\,,
\end{equation}
where $\sigma_a$ is -- as in the model proposed by \citet{zhou1995} -- the characteristic adhesive stress, and $\kappa$ is a parameter describing how rapidly adhesive contacts break under shear stress. In the adhesive, low-stress limit ($\sigma \!\to\! 0$), the maximum flowable fraction $\phi_{\rm{m}}$ equals $\phi_{\rm{alp}}$. For polydisperse mixtures such as the model dark paste, small cocoa particles fill the voids around larger sucrose crystals, so the system can remain flowable at higher $\phi$ even with adhesive contacts -- hence the larger low-$\sigma$ $\phi_{\rm{m}}$ for model dark pastes. This does not contradict the ``loose adhesive packing'' picture; it reflects packing \textit{geometry}, not stronger adhesion.

The fitting parameters are displayed in \cref{tab:RichardsParams}. It should be noticed that the errors related to the fitting parameters for the pre-refined dark pastes are higher than for the model pastes, because of a lack of low-stress data. Nevertheless, the fits -- shown as continuous lines in \cref{fig:phim_vs_sigma} -- show more clearly how the pre-refined paste switches from a behavior close to the one of the liquor at low stresses (showing a similar $\phi_{\rm{alp}}$) to a behavior close to the one of the model dark paste at high stresses (showing a similar $\phi_{\mu}$). It can also be observed that the fraction of contacts breaking under shear stress ($\kappa$) is highest (and equal to unity) for the model dark paste, which also shows the highest typical adhesive stress ($\sigma_a$). Our stress-dependent $\phi_{\rm{m}}(\sigma)$ map is consistent with the state-diagram picture advanced by \citet{blanco2019} for conched chocolates (although the samples in their study contain emulsifiers, which are not in the pastes investigated here), here made explicit by separating adhesive- and frictional-limit maximum fractions and by estimating the adhesive stress $\sigma_a$ and contact-breaking parameter $\kappa$.

\renewcommand{\tabcolsep}{5pt}
\begin{table}[!t]
    \centering
    \caption{Parameters obtained by fitting the data with the model proposed by \citet{richards2020} (\cref{eq:richards}).}
    \label{tab:RichardsParams}
    \begin{tabular}{|c|c|c|c|} \hline
    & pre-ref. dark & model dark & liquor \\ \hline
    $\phi_{\mu}$                    & $0.729 \pm 0.034$ & $0.733 \pm 0.002$ & $0.592 \pm 0.019$ \\ \hline
    $\phi_{\rm{alp}}$               & $0.404 \pm 0.277$ & $0.673 \pm 0.005$ & $0.400 \pm 0.040$ \\ \hline
    $\sigma_a \left[\rm{Pa}\right]$ & $1.85  \pm 4.40$  & $2.72  \pm 0.42$  & $2.65  \pm 1.69$  \\ \hline
    $\kappa$                        & $0.55  \pm 0.31$  & $1.00  \pm 0.14$  & $0.45  \pm 0.14$  \\ \hline
    \end{tabular}
\end{table}

Beyond extracting $\phi_{\rm{m}}(\sigma)$ via \cref{eq:MaronPierce,eq:richards}, we assessed the framework on \emph{entire families} of steady curves using the parameter sets per material from \cref{tab:RichardsParams}. As shown in \cref{app:modelflowcurves} (\cref{fig:full_flow_curves_richards_maronpierce}), model predictions overlap the data from \cref{fig:StressSweeps_PreRef,fig:StressSweeps_Model} with good fidelity; small deviations appear, especially at low stresses, as expected when adhesive contacts are most persistent.

In this framework, $\phi_\mu\equiv\lim_{\sigma\to\infty}\phi_{\rm{m}}(\sigma)$ is the friction-limit packing fraction (the maximum flowable solids when adhesive constraints are fully broken). At $\sigma=50\,\rm{Pa}$ we obtain (see \cref{tab:AlphasBetas}) $\phi_{\rm{m}}=0.675$ (pre-refined), $0.729$ (model dark), and $0.545$ (liquor), while the fitted limits are $\phi_\mu=0.729$, $0.733$, and $0.592$ (\cref{tab:RichardsParams}). Thus, the model dark paste is already very close to its friction limit at $50\,\rm{Pa}$, whereas the pre-refined dark and liquor pastes remain modestly below their limits.

Given the physical interpretation of $\phi_\mu$ and the estimated values of the characteristic adhesive stress $\sigma_a$, it is important to note that, within this framework, in the high-stress regime where $\sigma \gg \sigma_a$ and the system transitions from a lubricated to a frictional state \citep{morris2018}, the suspension becomes purely frictional and $\phi_\mu$ coincides with the jamming volume fraction $\phi_{\rm J}$ \citep{pednekar2017,singh2019}.

The comparatively large adhesive-limit $\phi_{\rm{alp}}$ for the model dark paste is consistent with its broad/bimodal particle-size distribution (see \cref{fig:PSDs}): small cocoa particles fill the voids around larger sucrose crystals, raising the packing efficiency even in the adhesive limit. This microstructural void-filling does not imply stronger adhesion; rather, it increases the attainable solid fraction before network breakup under stress.

Our stress-dependent flowability map $\phi_{\rm{m}}(\sigma)$ (\cref{fig:phim_vs_sigma}) and the separation of adhesive and frictional limits ($\phi_{\rm{alp}},\phi_\mu$) complement the state-diagram view of \citet{blanco2019}: their work shows how processing/lubrication moves a formulation across the diagram, while here we measure $\phi_{\rm{m}}(\sigma)$ and use it to predict full steady flow curves across $\phi$ with a single parameter set (see \cref{fig:full_flow_curves_richards_maronpierce} in \cref{app:modelflowcurves}).

\paragraph{Why friction matters}

In molten chocolate pastes, the solids are angular sucrose crystals and irregular cocoa particles dispersed in cocoa butter. At the fat levels and stresses studied here, particles can form direct (or asperity-mediated) contacts: sugar/cocoa surface asperities protrude through thin fat films, creating adhesive, friction-bearing bridges \citep{blanco2019}. Once such contacts percolate, the macroscopic resistance is set by two chocolate-specific parameters:
\begin{enumerate}[topsep=5pt,itemsep=5pt]
    \item \textbf{Contact friction set by surface coverage and roughness.} The thickness/continuity of cocoa-butter films determine whether contacts behave as dry/rough (higher $\mu$) or boundary-lubricated (lower $\mu$); higher $\mu$ promotes a lower friction-limit packing $\phi_\mu$ and a steeper viscosity rise as $\phi \to \phi_\mu$ \citep{blanco2019,boyer2011}.
    \item \textbf{Adhesion, $\sigma_a$, from fat-mediated and surface-energy effects.} Even with thin films present, attraction between sugar-cocoa asperities impose adhesive constraints that bear load at low stress. Increasing $\sigma$ breaks these adhesive contacts at a rate captured by $\kappa$, pushing the material toward lubrication-dominated flow; this is consistent with adhesive/cohesive yield-stress suspensions where attraction masks shear thickening and accentuates shear thinning \citep{bonn2017,buscall1987,mewis2011}.
\end{enumerate}

\noindent These chocolate-specific mechanisms map onto our observations: (i) positive bulk $F_N$ under rate control reflects the dilatancy of frictional/adhesive contact networks under confinement \citep{boyer2011}; (ii) at higher fat (lower $\phi$) and under stress control, contacts are relieved and the material accesses a boundary-lubricated branch (negative $F_N$, lower viscosity), consistent with contact release \citep{blanco2019}; (iii) the rise of $\phi_{\rm{m}}(\sigma)$ at low $\sigma$ records how adhesive contacts limit flow until stress breaks them \citep{bonn2017,mewis2011}; and -- as we will show in the next section -- (iv) in LAOS, the contact-stress fraction ranks liquor $>$ pre-refined $>$ model dark, matching stronger/more numerous contacts in cocoa-rich, narrower-PSD systems and more void-filling in the bimodal model-dark paste (cf.\ our steady-shear and LAOS analyses).

\subsection{Large Amplitude Oscillatory Shear (LAOS)}\label{sec:laos}

Oscillatory shear experiments were performed at a fixed frequency of $1\,\rm{Hz}$. Three strain amplitudes ($1\%$, $10\%$, and $100\%$) were applied, each for 100 cycles to ensure steady-state behavior and reproducibility. Since even the smallest amplitude exceeds the linear viscoelastic regime for these materials, these tests probe the nonlinear regime characteristic of LAOS.

\begin{figure*}[!h]
    \centering
    \includegraphics[width=\linewidth]{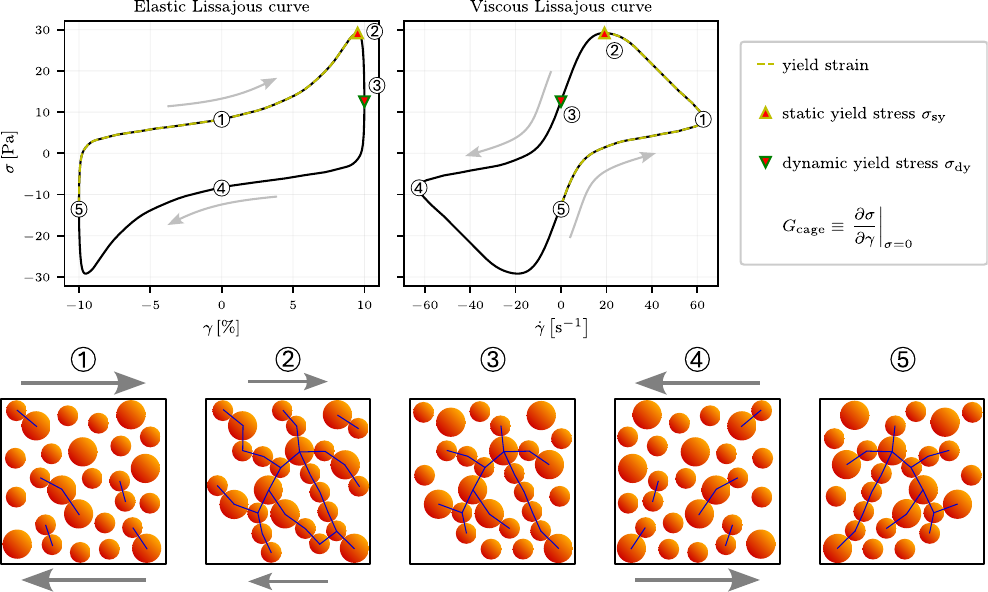}
    \vspace{-7pt}
    \caption{Elastic and viscous Lissajous curves for a representative LAOS cycle. Numbered points indicate key stages in the oscillation, as detailed in \cref{sec:laosanalysis}. Snapshots (fictional) of the microstructural rearrangements are proposed.}
    \label{fig:LAOS_exp}
\end{figure*}

Prior to each test, an oscillatory pre-shear was applied, with strain amplitude decreasing from 50\% to 0.01\% at $2\,\rm{Hz}$ over 10 minutes. This conditioning effectively erases residual structural memory induced by sample loading and minimizes anisotropy or preferential microstructural alignment that could bias measurements. Such preconditioning is especially critical near or beyond the jamming transition (in our pre-refined samples, this is at 24 and 20\,\rm{wt}\% fat, respectively), where the microstructure is highly sensitive to shear history. All tests were conducted at 40{\textdegree}C. 

We first outline the methodology and fundamental concepts underpinning LAOS before presenting experimental results.

\paragraph{Bridge to steady shear}

LAOS complements the steady-shear analysis organized by the stress-dependent maximum flowable fraction $\phi_{\rm{m}}(\sigma)$, which delineates the yield loci in $(\phi,\sigma)$. Steady tests locate each composition relative to these loci, while LAOS resolves how contact networks build and break within a cycle as the material is driven across this boundary. Accordingly, trends in LAOS static/dynamic yield stresses, normalized yield strain, and the cage modulus $G_{\text{cage}}$ track distance to $\phi_{\rm{m}}(\sigma)$; and the contact-stress decomposition at $\gamma_0=100\%$ reproduces the same contact- vs. lubrication-dominated ordering inferred from steady sweeps.

\subsubsection{LAOS Analysis}\label{sec:laosanalysis}

During LAOS, the applied strain is sinusoidal, $\gamma(t) = \gamma_0 \sin(\omega t)$, where $\gamma_0$ is the strain amplitude and $\omega$ the frequency. \Cref{fig:LAOS_exp} illustrates a typical chocolate paste stress response over one cycle, with key points numbered as follows:
\begin{enumerate}[topsep=0pt]
    \item Strain is zero; shear rate is at its maximum.
    \item Strain increases while shear rate decreases; the contact network becomes denser in connections, causing a (positive) stress peak.
    \item Shear rate crosses zero and reverses; stress rapidly declines as contacts open.
    \item Strain increases in the opposite direction and crosses zero, where shear rate is at its (negative) maximum; microstructure rearranges and the stress increases in magnitude.
    \item Shear rate decreases and reaches zero; shear strain is at its (negative) maximum.
\end{enumerate}
After point 5, shear rate increases and strain decreases going back to zero initiating a new cycle. This sequence illustrates the dynamic coupling between deformation, shear rate, and microstructural response during oscillation.

Notably, for the chocolate pastes studied here, as shown in the example, the peak stress (point 2) often precedes the strain maximum (point 3), at which the shear rate reverses. At the reversal point (3), stress remains positive despite the instantaneous shear rate changing sign, reflecting a stress lag caused by microstructural relaxation and contact breakage.

\cref{fig:LAOS_exp} also presents a schematic of the proposed microstructural rearrangement during the LAOS cycle: at zero strain (point 1), some contacts may already exist; then, strain increases and the contact network becomes denser in connections, yielding a stress peak (the static yield stress, point 2); at zero shear rate and maximum strain (point 3), some contacts relax and open, but some  contacts, perhaps including adhesive ones, persist (resulting in the dynamic yield stress), showing a positive stress; when shear strain crosses zero and shear rate reaches its maximum in the reversed direction (point 4), the microstructure is  expected to be a mirror image of that at point 1; the shear rate will then decrease and reach zero, to a situation analogous to the one at point 3 but in the reversed direction; finally, shear rate increases and strain goes back to zero initiating a new cycle.

The nonlinear stress response $\sigma(t)$ can be decomposed into elastic and viscous components via Fourier Transform (FT) rheology \citep{ewoldt2008,cho2005,choi2019}:
\begin{equation}
\begin{split}
    \sigma(t) = \gamma_0 \sum_{n\ \rm{odd}} \left[\right. &G'_n\left(\omega,\gamma_0\right)\, \sin(n \omega t) \\ +\ &G''_n\left(\omega,\gamma_0\right)\, \cos(n \omega t) \left.\right]\,.
\end{split}
\end{equation}
Higher odd harmonics ($n=3,5,\dots$) quantify nonlinearities beyond the fundamental oscillation, detecting deviations from linear viscoelasticity.

The sequence of physical processes (SPP) framework \citep{rogers2011,rogers2012,rogers2012a,lee2017,rogers2017} provides an effective tool for analyzing LAOS responses in complex food materials \citep{joyner2021,duvarci2017,erturk2023,vandervaart2013}. A key parameter derived from this analysis is the cage modulus:
\begin{equation}
    G_{\rm cage} \equiv \left. \frac{d\sigma}{d\gamma} \right|_{\sigma=0}\,,
\end{equation}
which characterizes the local stiffness at zero global stress. This parameter is valuable for understanding yielding in structured materials such as chocolate paste, where microstructural cages constrain particle motion prior to flow \citep{sandoval2022}.

Additional metrics include yield strain and static and dynamic yield stresses. Yield strain quantifies deformation between the last strain reversal and the maximum stress within a cycle. Purely elastic materials exhibit yield strain near $2\gamma_0$, while purely viscous materials yield near $\gamma_0$. The static yield stress corresponds to the threshold for flow initiation (maximum stress, i.e., stress at yield strain), while the dynamic yield stress corresponds to the stress at maximum strain (i.e., at $\dot{\gamma} = 0$), below which flow ceases.

\begin{figure*}[!h]
    \centering
    \includegraphics[width=\linewidth]{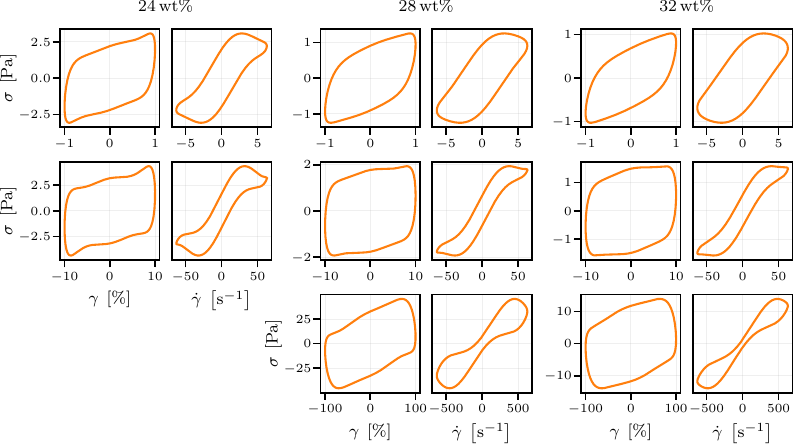}
    \caption{Elastic and viscous Lissajous curves for pre-refined pastes at a selection of fat contents, at strain amplitudes of (top to bottom) 1\%, 10\%, and 100\% and at $1\,\rm{Hz}$.}
    \label{fig:LAOS_Lissajous_PreRef}
\end{figure*}

\subsubsection{LAOS Results}

Pre-refined dark chocolate pastes were tested at $1\,\rm{Hz}$ with strain amplitudes $\gamma_0 = 1$, 10 and 100\% and fat contents of $20$, $24$, $28$, $32$, and $40\,\rm{wt}\%$. Representative Lissajous curves are shown in \cref{fig:LAOS_Lissajous_PreRef}. The shapes vary strongly with solid fraction and strain amplitude, reflecting complex microstructural rearrangements and nonlinear rheology. LAOS enables probing the material beyond jamming under controlled conditions: samples at 20 and 24$\,\rm{wt}\%$ fat, which do not flow under steady shear, yielded measurable responses at low strains but fractured at $\gamma_0=100\%$, and hence the lack of a Lissajous figure for this condition at lower left in \cref{fig:LAOS_Lissajous_PreRef}.

\begin{figure}[!t]
    \centering
    \includegraphics[width=\linewidth]{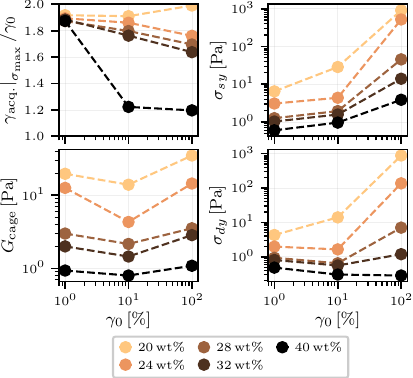}
    \caption{Pre-refined pastes: normalized yield strain (top left), cage modulus (bottom left), static yield stress (top right), and dynamic yield stress (bottom right) vs. strain amplitude, varying fat content. Lines are a visual guide.}
    \label{fig:LAOS_spp_PreRef}
\end{figure}

\begin{figure}[!t]
    \centering
    \includegraphics[width=\linewidth]{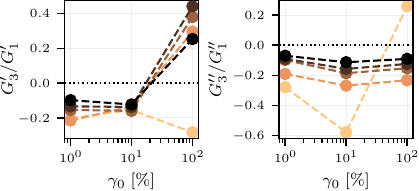}
    \caption{Pre-refined pastes: harmonic ratios $G'_3/G'_1$ (left) and $G''_3/G''_1$ (right) vs. strain amplitude, varying fat content. Lines are a visual guide. Same legend as in \cref{fig:LAOS_spp_PreRef}.}
    \label{fig:LAOS_harmonic_ratios_PreRef}
\end{figure}

The first nonlinearity is observed directly in the Lissajous curves: for all fat contents, the stress has a much sharper rise from amplitude 10\% to 100\% relative to 1\% to 10\%.

\Cref{fig:LAOS_spp_PreRef} displays normalized yield strain, cage modulus, and static and dynamic yield stresses as functions of strain amplitude for different fat contents. Yield strain is near $2\gamma_0$ at low amplitude, consistent with elastic behavior. Increasing strain amplitude shifts the response toward viscous behavior, especially at high fat content (low solid volume fraction), with the $40\,\rm{wt}\%$ paste approaching approximately $1.2\gamma_0$ at $\gamma_0=100\%$. The cage modulus generally decreases with fat content but increases with strain amplitude, indicating strain-dependent stiffening. Static and dynamic yield stresses follow similar trends, although the dynamic yield stress slightly decreases with amplitude at the highest fat contents ($32$ and $40\,\rm{wt}\%$).

In addition to the SPP metrics, we quantified the standard LAOS harmonic ratios $G'_3/G'_1$ and $G''_3/G''_1$ from odd-harmonic fits of $\sigma(t)$. As shown in \cref{fig:LAOS_harmonic_ratios_PreRef}, $G'_3/G'_1$, increases with $\gamma_0$ and flips sign switching from intra-cycle strain-stiffening to intra-cycle strain-softening at high amplitudes, while $G''_3/G''_1$ remains negative with $\gamma_0$ (intra-cycle viscous shear-thinning); both ratios have larger magnitude at lower fat (higher $\phi$) and are reduced at higher fat (lower $\phi$), consistent with stronger vs.\ weaker contact-mediated nonlinearity, respectively. For both quantities, the lowest-fat-content paste shows odd behavior at high amplitudes.

\begin{figure*}[!h]
    \centering
    \includegraphics[width=\linewidth]{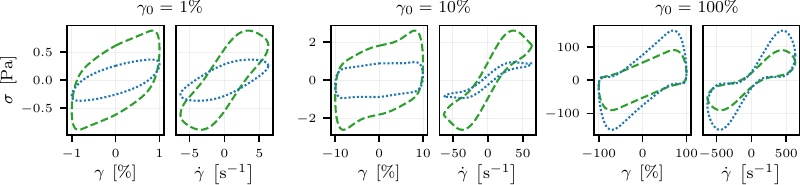}
    \caption{Elastic and viscous Lissajous curves for model dark (dashed green, $\phi = 0.71$) and liquor (dotted blue, $\phi = 0.53$) pastes, at strain amplitudes of 1\%, 10\%, and 100\% (left to right) and at $1\,\rm{Hz}$.}
    \label{fig:LAOS_Lissajous_Model}
\end{figure*}

For model dark chocolate and liquor pastes, experiments focused near the maximum flowable volume fractions at $\sigma=50\,\rm{Pa}$: $\phi=0.71$ (model dark) and $\phi=0.53$ (liquor). Strain amplitudes $\gamma_0 = \{1, 10, 100\}\%$ were applied at $1\,\rm{Hz}$. Lissajous curves are shown in \cref{fig:LAOS_Lissajous_Model}. At $\gamma_0=100\%$, the liquor paste exhibits higher stress than the dark paste, reversing the relationship seen at lower strain amplitudes. Model pastes show the same sharp stress increase observed for pre-refined pastes from amplitude 10\% to 100\%.

\begin{figure}[!t]
    \centering
    \includegraphics[width=\linewidth]{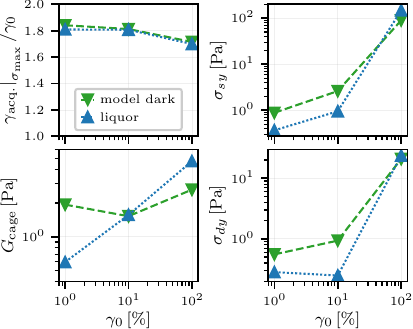}
    \caption{Model dark and liquor pastes: normalized yield strain (top left), cage modulus (bottom left), static yield stress (top right), and dynamic yield stress (bottom right) vs. strain amplitude. Lines are a visual guide.}
    \label{fig:LAOS_spp_Model}
\end{figure}

\begin{figure}[!t]
    \centering
    \includegraphics[width=\linewidth]{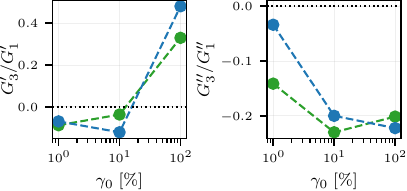}
    \caption{Model pastes: harmonic ratios $G'_3/G'_1$ (left) and $G''_3/G''_1$ (right) vs. strain amplitude for different fat contents. Lines are a visual guide. Same legend as in \cref{fig:LAOS_spp_Model}.}
    \label{fig:LAOS_harmonic_ratios_Model}
\end{figure}

Yield strain remains close to $2\gamma_0$ for both pastes across all amplitudes, with minor decreases. The cage modulus increases with strain amplitude, indicating non-monotonic strain-dependent stiffness. Both static and dynamic yield stresses increase monotonically with strain amplitude (see \cref{fig:LAOS_spp_Model}). It should be noted that, albeit at $\gamma_0 = 100\%$ the liquor paste Lissajous curves enclose the ones for the model dark paste (the opposite happens at lower amplitudes), the static yield stress (i.e., the peak stress) of both pastes is similar.

Applying the same analysis to the model pastes shows the same qualitative amplitude trends -- $G'_3/G'_1$ rises and $G''_3/G''_1$ becomes more negative with increasing $\gamma_0$ -- but with systematically smaller magnitudes at a given $\phi$ and $\gamma_0$ than in pre-refined pastes. This attenuation of nonlinearity is consistent with the model system's bimodal/void-filling microstructure and its more lubrication-dominated response at comparable $\phi$, in line with the steady-shear $\phi_{\rm{m}}(\sigma)$ map and the LAOS contact–stress ordering. Taken together, the harmonic ratios corroborate the SPP interpretation and mirror the approach to the stress-dependent flow limit in the $\phi_{\rm{m}}(\sigma)$ map: pre-refined formulations (more contact-dominated) exhibit larger $G'_3/G'_1$ and more negative $G''_3/G''_1$ than model pastes at the same $\gamma_0$ and $\phi$.

\subsubsection{Viscosity and Relaxation Strain}

To probe microstructural relaxation, normalized relative viscosity immediately after strain reversal and relaxation strain were analyzed for all pastes.

\begin{figure}[!t]
    \centering
    \includegraphics[width=\linewidth]{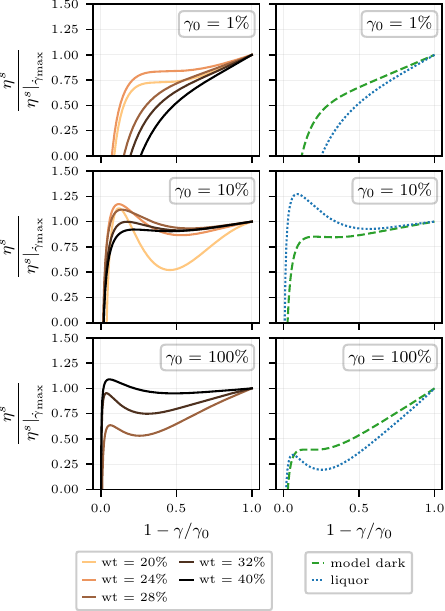}
    \caption{Normalized relative viscosity vs. inverted normalized strain ($1-\gamma/\gamma_0$) immediately after reversal at strain amplitudes of (top) 1\%, (center) 10\%, and (bottom) 100\%, for various fat contents and (left) pre-refined pastes (same legend as in \cref{fig:LAOS_relaxStrain}) and (right) model dark (dashed green, $\phi = 0.71$) and liquor (dotted blue, $\phi = 0.53$) pastes.}
    \label{fig:LAOS_etaS}
\end{figure}

\Cref{fig:LAOS_etaS} depicts normalized relative viscosity after shear reversal as a function of the normalized strain for all pastes at various solid fractions and strain amplitudes; this is plotted in inverted form $1-\gamma/\gamma_0$. The relative viscosity has been normalized by its value at maximum shear rate -- corresponding to $\gamma = 0$ -- so that all the plotted curves show recovery to 1 after reversal. Note that the viscosity defined in terms of instantaneous values of shear stress and shear rate can be negative, as seen in the plot. At low amplitude, viscosity recovers slowly to pre-reversal levels, indicating slow microstructural relaxation. At intermediate amplitude, recovery is faster and accompanied by transient viscosity overshoots for certain fat contents. At high amplitude (noting fracturing at this amplitude for $20$ and $24\,\rm{wt}\%$ pre-refined samples), viscosity rapidly rises post-reversal, then exhibits diverse behaviors: plateau and slow recovery, overshoot followed by undershoot and slow recovery, or overshoot with full recovery.

\begin{figure}[!t]
    \centering
    \includegraphics[width=\linewidth]{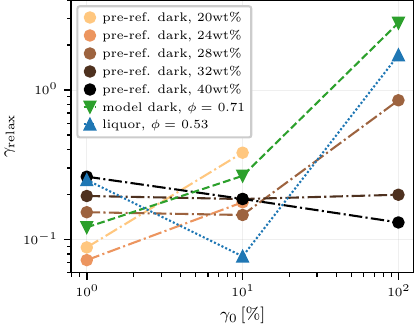}
    \caption{Relaxation strain vs. strain amplitude. Lines are a visual guide.}
    \label{fig:LAOS_relaxStrain}
\end{figure}

The relaxation strain -- i.e., the strain required post-reversal for viscosity to regain positive correlation with shear rate -- is shown in \cref{fig:LAOS_relaxStrain}. Samples near or above the jamming transition show increasing relaxation strain with increasing amplitude, except for the liquor paste at $\gamma_0=10\%$ that is an outlier, whereas $32$ and $40\,\rm{wt}\%$ pre-refined dark pastes exhibit slightly decreasing relaxation strain with amplitude, suggesting distinct microstructural relaxation.

\subsubsection{Contact Stress Decomposition}

To elucidate underlying mechanisms, contact stress decomposition (CD) -- a method proposed by \citet{yanjie2024} -- was applied to separate hydrodynamic and contact stress contributions.

Following \citet{yanjie2024} and based on the shear reversal analysis conducted by \citet{peters2016}, the hydrodynamic viscosity $\eta_{H}$ is determined from the minimum of the tangential viscosity $\eta_t = \partial \sigma / \partial \dot{\gamma}$ after the reversal point, while the pre-reversal plateau represents $\eta_H + \eta_C$, from which one can then easily retrieve $\eta_{C}$. Thus,
\begin{equation}
    \sigma = \sigma_H + \sigma_C = (\eta_H + \eta_C) \dot{\gamma}\,.
\end{equation}

\begin{figure}[!t]
    \centering
    \includegraphics[width=\linewidth]{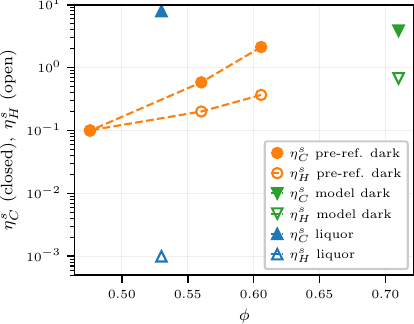}
    \caption{Contact (closed symbol) and hydrodynamic (open symbols) relative viscosities at $\gamma_0=100\%$, following \citet{yanjie2024}. Trends mirror the contact– vs.\ lubrication–dominated ordering inferred from steady sweeps, consistent with the $\phi_{\rm{m}}(\sigma)$ framework.}
    \label{fig:LAOS_CSD}
\end{figure}

However, in adhesive materials, the viscosity minimum after reversal may not correspond to the hydrodynamic viscosity because of adhesive contacts that may survive the change of shear direction. Hence, we decided to perform CD at the highest strain amplitude $\gamma_0=100\%$ only, as all contact would break with such high strain amplitude. \Cref{fig:LAOS_CSD} shows that both relative viscosities increase with volume fraction $\phi$ for pre-refined pastes. For comparable volume fractions, the liquor paste exhibits higher contact viscosity and lower hydrodynamic viscosity with respect to the pre-refined paste.

Although this last measure is performed at high amplitude only, it confirms once again the central role of frictional and adhesive contacts in the stress transmission of chocolate pastes.

\section{Conclusions}\label{sec:conclusions}

This work has presented a comprehensive rheological characterization of pre-refined dark chocolate pastes, combining industrially relevant pre-refined formulations with controlled model systems. By applying both steady shear and large-amplitude oscillatory shear (LAOS) protocols, we have shown how fat content (i.e., solid volume fraction) and particle-scale interactions govern yielding, flow, and microstructural dynamics.

In steady shear, stress- and rate-controlled sweeps consistently reveal yield stress behavior accompanied by shear thinning. Pre-refined pastes display a pronounced decrease in yield stress with fat content and a transition of normal stresses from positive to negative, indicating a shift in the balance of interparticle forces. Model systems exhibit richer behavior: shear thinning with weak thickening in dark chocolate pastes depending on solid fraction, and persistent shear thinning in cocoa liquor pastes. Stress sweeps also reveal hysteresis near yielding, highlighting the role of structural memory.

Analysis of the viscosity data through the Maron-Pierce model provides stress-dependent maximum flowable fractions $\phi_{\rm m}(\sigma)$, which delineate yield loci in the $(\phi,\sigma)$ plane. Pre-refined pastes show a strong stress dependence of $\phi_{\rm m}$, consistent with highly frictional and adhesive contacts, while model systems exhibit weaker variations. Fitting $\phi_{\rm m}(\sigma)$ with the \citet{richards2020} framework situates chocolate pastes within the broader context of constraint rheology, quantifying parameters such as the adhesive stress scale and contact-breaking rate.

Nonlinear oscillatory tests further extend the analysis to regimes near and beyond jamming. Lissajous curves, cage modulus, and yield strain reveal systematic transitions from elastic to viscous response with increasing strain amplitude and fat content, while relaxation strain measurements expose distinct recovery pathways close to the jamming threshold. Contact stress decomposition under LAOS confirms the central role of frictional and adhesive contacts in stress transmission, particularly in pre-refined pastes.

Altogether, this study establishes chocolate pastes as dense, adhesive suspensions whose rheology is most strongly impacted by the solid fraction, but whose detailed behavior is governed by the interplay of particle size distribution, adhesion, and friction. The consistency of findings across steady shear, stress sweeps, and LAOS offers benchmarks for constitutive modeling, while the stress-dependent $\phi_{\rm m}$ and contact stress decomposition provide direct input for predictive frameworks. This integrated picture advances the rheology of dense adhesive suspensions and informs processing of industrial chocolate formulations.

Looking ahead, several directions emerge. Systematic constitutive modeling that unifies stress-dependent jamming with nonlinear oscillatory responses could build predictive frameworks linking microstructural dynamics to macroscopic rheology. Extending the methodology to other food suspensions and controlled systems would test the generality of the observed stress- and strain-dependent transitions. Finally, coupling rheometry with in situ structural probes -- such as scattering or imaging -- would directly connect microstructural evolution to measured constitutive behavior. These perspectives would not only refine our understanding of chocolate pastes but also advance the broader physics of dense, jammed, and adhesive suspensions.

\section*{Acknowledgements}
This work was supported by Mondel\={e}z International. MO and JFM were partially supported by NSF CBET-2228680.

\section*{Declarations}
\textbf{Conflict of interest} The authors have no conflicts to disclose.

\appendix
\crefalias{section}{appendix}

\section{Reproducibility}\label{app:reproducibility}

For the steady-shear data, three independent loadings were performed for each sample and composition. In \cref{fig:etaS_vs_phi_stress50Pa_PreRef,fig:etaS_vs_phi_stress50Pa_Model}, each point and its error bar represent, respectively, the mean and the standard deviation (SD) of the steady viscosity across independent loadings.

\begin{figure}[!h]
    \centering
    \vspace{-12pt}
    \includegraphics[width=\linewidth]{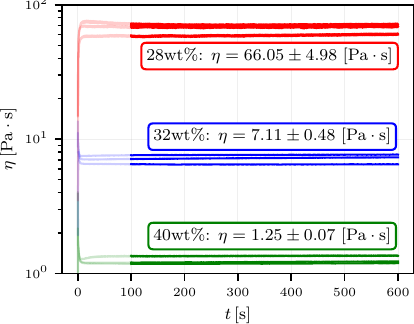}
    \vspace{-12pt}
    \caption{Raw viscosity vs.\ time at $\sigma=50\,\rm{Pa}$ for representative pre-refined samples at 28, 32, and 40\,wt\% fat contents: three independent loadings per composition. The transparent initial window indicates the first $100\,\rm{s}$ discarded before steady-state averaging. Insets report the mean $\pm$ SD over the steady window for each composition. The good agreement demonstrate high reproducibility and explain the small error bars shown in \cref{fig:etaS_vs_phi_stress50Pa_PreRef,fig:etaS_vs_phi_stress50Pa_Model}.}
    \label{fig:reproducibility_PreRef}
\end{figure}

\vspace{-20pt}
\section{Full-curve validation of the constraint framework}\label{app:modelflowcurves}

\begin{figure}[!t]
    \centering
    \includegraphics[width=\linewidth]{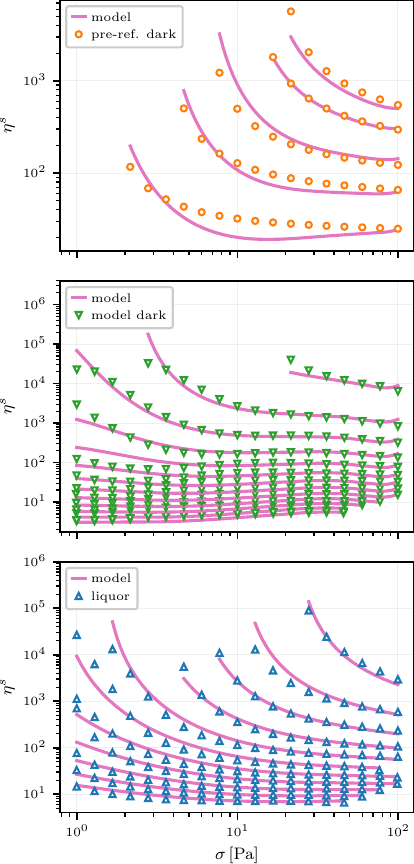}
    \vspace{-12pt}
    \caption{Validation on full families of steady curves using a single parameter set per material.
    For each system, symbols show relative viscosity $\eta^s$ vs. stress $\sigma$ at several volume fractions $\phi$; lines are predictions from the model proposed by \citet{richards2020} given by the combination of \cref{eq:MaronPierce,eq:richards} and parameters from \cref{tab:RichardsParams}.}
    \label{fig:full_flow_curves_richards_maronpierce}
    \vspace{-8pt}
\end{figure}

\cref{fig:full_flow_curves_richards_maronpierce} tests the constraint framework \citep{richards2020} on \emph{entire families} of flow curves using a \emph{single} parameter set per material (\cref{tab:RichardsParams}). For each material and several solid fractions $\phi$ spanning the experimental range, we show the relative viscosities with predictions obtained from \cref{eq:MaronPierce,eq:richards}.

Each panel corresponds to a fixed material; symbols are experimental steady data at fixed $\phi$ and varying $\sigma$ (multiple $\phi$ per panel), and solid lines are the corresponding model predictions with the \emph{single} parameter set from \cref{tab:RichardsParams}. Agreement across decades in $\sigma$ indicates that (i) the stress-dependent flowability map $\phi_{\rm{m}}(\sigma)$ extracted from \cref{eq:MaronPierce,eq:richards} and (ii) the Maron-Pierce divergence jointly capture the observed families of flow curves. Small discrepancies appearing at the lowest stresses are expected: in that regime adhesive contacts are most persistent and time-to-steady can be longest, so modest deviations are not diagnostic of model failure.

The good agreement demonstrate that a single, independently fitted set of $[\phi_{\mathrm{alp}},\phi_\mu,\sigma_a,\kappa,\beta,\alpha(\sigma)]$ (noting that the parameter $\alpha(\sigma)$, i.e., the prefactor in \cref{eq:MaronPierce}, is stress-dependent -- its values across the stresses investigated here are shown in the inset of \cref{fig:phim_vs_sigma}) per material suffices to predict steady flow curves across compositions. This supports the physical interpretation adopted in the main text: $\phi_{\rm{m}}(\sigma)$ increases monotonically from an adhesive limit toward a frictional limit, and the viscosity families follow accordingly via the Maron-Pierce form.

Predictions are for steady-state responses; they are not intended to capture transient startup or history effects. All curves use the same averaging protocol and baseline viscosity (cocoa butter) as in the main figures, ensuring consistency between main-text and this analysis. For the model paste, predictions use the increasing-stress branch consistently with the $\phi_{\rm{m}}(\sigma)$ extraction discussed in the main text.

\printbibliography

\end{document}